\documentclass[a4paper]{jpconf}

\usepackage{hyperref}
\usepackage[dvips]{epsfig}

\begin{document}

\title{Role of transverse displacements for a quantized-velocity state of
  the lubricant}

\author{Ivano Eligio Castelli$^1$, Nicola Manini$^{1,2}$,
    Rosario Capozza$^3$, Andrea Vanossi$^3$,
        Giuseppe E. Santoro$^{2,4}$, and Erio Tosatti$^{2,4}$}
\address{
$^1$Dipartimento di Fisica and CNR-INFM, Universit\`a di Milano, Via
Celoria 16, 20133 Milano, Italy
}
\address{
$^2$International School for Advanced Studies (SISSA)
and CNR-INFM Democritos National Simulation Center, Via Beirut 2-4, I-34014
Trieste, Italy
}
\address{
$^3$CNR-INFM National Research Center S3 and Department of Physics, \\
University of Modena and Reggio Emilia, Via Campi 213/A, 41100 Modena,
Italy
}
\address{
$^4$International Centre for Theoretical Physics (ICTP), P.O.Box 586,
I-34014 Trieste, Italy
}

\date{January 31, 2008}

\begin{abstract}
Within the idealized scheme of a 1-dimensional Frenkel-Kontorova-like
model, a special ``quantized'' sliding state was found for a solid lubricant
confined between two periodic layers~\cite{Vanossi06}.
This state, characterized by a nontrivial geometrically fixed ratio of the
mean lubricant drift velocity $\langle v_{\rm cm}\rangle$ and the
externally imposed translational velocity $v_{\rm ext}$, was understood as
due to the kinks (or solitons), formed by the lubricant due to
incommensuracy with one of the substrates, pinning to the other sliding
substrate.
A quantized sliding state of the same nature is demonstrated here for a
substantially less idealized 2-dimensional model, where atoms are allowed
to move perpendicularly to the sliding direction and interact via
Lennard-Jones potentials.
Clear evidence for quantized sliding at finite temperature is
provided, even with a confined solid lubricant composed of
multiple (up to 6) lubricant layers.
Characteristic backward lubricant motion produced by the presence of
``anti-kinks'' is also shown in this more realistic context.
\end{abstract}



\section{Introduction}
The problem of lubricated friction is a fascinating one, both from
the fundamental point of view and for applications. Lubricants
range from thick fluid layers to few or even single mono-layers,
often in a solid or quasi-solid phase (boundary lubrication).
In the present work, we address the effects of lattice parameter mismatch
of the solid boundary lubricant and the two confining crystalline surfaces.
In general, perfect inter-atomic matching tends to produce locking, while
sliding is always favored by ``defective'' lines (misfit dislocations),
which can be introduced precisely by incommensuration of the lubricant and
the sliding substrate lattice parameters.
In our 3-length scale slider-lubricant-slider confined geometry, this
lattice mismatch may give rise to a very special ``quantized'' sliding
regime, where the mean lubricant sliding velocity, is fixed to an exact
fraction of the relative substrate sliding velocity.
This velocity fraction, in turn, is a simple function of the
lubricant ``coverage'' with respect to the less mismatched of the
two substrate surfaces \cite{Vanossi06,Manini07PRE}.
This special sliding mode was discovered and analyzed in detail in a very
idealized 1-dimensional (1D) Frenkel-Kontorova (FK)-like
model~\cite{Vanossi06}: the plateau mechanism was interpreted in terms of
solitons, or kinks (the 1D version of misfit dislocations), being produced
by the mismatch of the lubricant periodicity to that of the more
commensurate substrate, with these kinks being rigidly dragged by the
other, more mismatched, substrate.

In the present work, we investigate the presence of similar
velocity plateaus associated to solitonic mechanisms in a more
realistic geometry: a 2-dimensional (2D) $x-z$ model of
Lennard-Jones (LJ) solid lubricant.

\section{The 2D model}\label{model:sec}

\begin{figure}
\centerline{
\epsfig{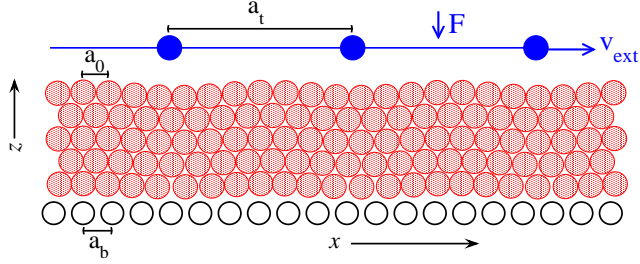}
}
\caption{\label{model:fig}
A sketch of the model with the rigid top (solid circles) and bottom (open)
layers (of lattice spacing $a_{\rm t}$ and $a_{\rm b}$ respectively), the
former moving at externally imposed $x$-velocity $v_{\rm ext}$.
One or more lubricant layers (shadowed) of rest equilibrium spacing $a_{\rm
0}$ are confined in between.
}
\end{figure}

We represent the sliding crystalline substrates by two rows of
equally-spaced ``atoms''.
Between these two rigid layers, we insert $N_{\rm p}$ identical lubricant
atoms, organized in $N_{\rm layer}$ layers (see Fig.~\ref{model:fig} where
$N_{\rm layer}=5$).
While the mutual position of top and bottom substrate atoms are fixed,
the lubricant atoms move under the action of pairwise LJ potentials
\begin{equation}\label{LJpotential}
\Phi_a(r)=\epsilon_a
\left[\left(\frac{\sigma_a}{r}\right)^{12}
-2\left(\frac{\sigma_a}{r}\right)^6\right]
\end{equation}
describing the mutual interactions between them, and with the substrate
atoms as well.
To avoid long-range tails, we set a cutoff radius at $r=r_{\rm c}=
2.5\,\sigma_a$, where $\Phi_a\left(r_{\rm c} \right) \simeq -8.2\,\cdot
10^{-3}\,\epsilon_a$.

For the two substrates and the lubricant we assume three different kinds of
atoms, and characterize their mutual interactions as truncated-LJ
potentials ($\Phi_{\rm bp}$, $\Phi_{\rm pp}$ and $\Phi_{\rm tp}$ refer to
interaction energies for the bottom-lubricant, lubricant-lubricant, and
top-lubricant interactions, respectively) with the following LJ radii
$\sigma_a$
\begin{equation}\label{sigma}
\sigma_{\rm tp}=a_{\rm t}\,,\qquad
\sigma_{\rm bp}=a_{\rm b}\,,\quad
{\rm and} \
\sigma_{\rm pp}=a_{\rm 0} \,,
\end{equation}
which, for simplicity, are set to coincide with the fixed spacings $a_{\rm
t}$ and $a_{\rm b}$ between neighboring substrate atoms, and the average
$x$-separation $a_{\rm 0}$ of two neighboring lubricant atoms,
respectively.
This restriction is only a matter of convenience, and is not essential to
the physics we are describing. 
The choice of slightly different values of $\sigma_{\rm tp}$ and
$\sigma_{\rm bp}$ does not affect the lubricant to substrate density
ratios, which are the crucial ingredient driving the ``quantized'' sliding
state we address here: accordingly very similar results are observed.
If however the LJ radii were taken much larger or smaller than the
corresponding lattice parameters, then undesired phenomena could occur,
such as lubricant atoms squeezing in between the substrate layers and
escaping confinement altogether.
The three different periodicities $a_{\rm t}$, $a_{\rm 0}$ and $a_{\rm
b}$ define two independent dimensionless ratios:
\begin{equation}\label{ratiotb}
\lambda_{\rm t}=\frac{a_{\rm t}}{a_{\rm 0}}\,,\qquad
\lambda_{\rm b}=\frac{a_{\rm b}}{a_{\rm 0}}\,,
\end{equation}
the latter of which we take closer to unity, $\max(\lambda_{\rm
  b},\lambda_{\rm b}^{-1})<\lambda_{\rm t}$, so that the lubricant is
closer in registry to the bottom substrate than to the top.

For simplicity, we fix the same LJ interaction energy $\epsilon_{\rm tp}
=\epsilon_{\rm pp} =\epsilon_{\rm bp} =\epsilon$ for all pairwise coupling
terms. We also assume the same mass $m$ for all particles.
We take $\epsilon$, $a_{\rm 0}$, and $m$ as energy, length, and mass units.
This choice defines a set of ``natural'' model units for all physical
quantities, for instance velocities are measured in units of
$\epsilon^{1/2} \, m^{-1/2}$.
In the following, all mechanical quantities are expressed implicitly in the
respective model units.

The interaction with the other lubricant and sliders' particles produces a
total force of the $j$-th lubricant particle
\begin{eqnarray}\label{Fj}
\vec F_j&=& -
\sum_{i=1}^{N_{\rm t}}\frac{\partial}{\partial \vec r_j}
\Phi_{\rm tp}\!\left(|\vec r_j-\vec r_{{\rm t}\,i}|\right) +
\\ \nonumber
&&
- \sum_{j'=1 \atop j'\ne j}^{N_{\rm p}}
\frac{\partial}{\partial \vec r_j}
\Phi_{\rm pp}\!\left(|\vec r_j-\vec r_{j'}|\right)
-\sum_{i=1}^{N_{\rm b}}
\frac{\partial}{\partial \vec r_j}
\Phi_{\rm bp}\!\left(|\vec r_j-\vec r_{{\rm b}\,i}|\right)
,
\end{eqnarray}
where $\vec r_{{\rm t}\,i}$ and $\vec r_{{\rm b}\,i}$ are the positions of
the $N_{\rm t}$ top and $N_{\rm b}$ bottom atoms.
By convention, we select the frame of reference where the bottom layer is
immobile.
The top layer moves rigidly at a fixed horizontal velocity $v_{\rm ext}$,
and can also move vertically (its inertia equals the total mass $N_{\rm
  t}m$ of its atoms) under the joint action of the external vertical force
$-F$ applied to each particle in that layer plus that due to the interaction
with the particles in the lubricant layer:
\begin{equation}\label{xztop}
r_{{\rm t}\,i\,x}(t)= i\,a_{\rm t}+v_{\rm ext}\,t
\,,\qquad
r_{{\rm t}\,i\,z}(t)= r_{{\rm t}\,z}(t)
\,,
\end{equation}
where the equation governing $r_{{\rm t}\,z}$ is
\begin{eqnarray}\label{zztop}
N_{\rm t}m \,\ddot r_{{\rm t}\,z} &=&\!-\! \sum_{i=1}^{N_{\rm
t}}\sum_{j=1}^{N_{\rm p}} \frac{\partial \Phi_{\rm tp}}{\partial
r_{{\rm t}\,i\,z}} \!\left(|\vec r_{{\rm t}\,i}-\vec
r_j|\right)-\!N_{\rm t}F \,.
\end{eqnarray}

To simulate finite temperature in this driven model, we use a
standard implementation of the Nos\'e-Hoover thermostat chain
\cite{nose-hoover,Martyna92}, rescaling particle velocities with
respect to the instantaneous lubricant center of mass (CM)
velocity $v_{\rm cm}$.
The Nos\'e-Hoover chain method is described by the following equations
\cite{Martyna92}:
\begin{eqnarray} \label{nose-eq}
m\,\ddot {\vec r}_j&=& \vec F_j-\xi_1 m\,(\dot{\vec r}_j-\vec
v_{\rm cm})
\,,\\
\dot \xi_1&=&\frac{1}{Q_1} \left(\sum_{j=1}^{N_p} \left|\dot{\vec
r}_j-\vec v_{\rm cm}\right|^2 -gK_BT\right)-\xi_1\xi_2
\,,\\
\dot \xi_i&=&\frac{1}{Q_i}
\left(Q_{i-1}\xi^2_{i-1}-K_BT\right)-\xi_i\xi_{i+1}
\,,\\
\dot {\xi_M}&=&\frac{1}{Q_M} \left(Q_{M-1}\xi^2_{M-1}-K_BT\right)
\,.
\end{eqnarray}
The thermostat chain acts equally on all lubricant
particles $j=1,... N_{\rm p}$.
The $M=3$ thermostats are characterized by the effective ``mass''
parameters $Q_1=N_{\rm p}$, $Q_2=Q_3=1$ ; the coefficient
$g=2\left( N_{\rm p}-1\right)$ fixes the correct equipartition;
the auxiliary variables $\xi_i$ ($i=1,...\, M$) keep the kinetic
energy of the lubricant close to its classical value $N_{\rm
p}K_BT$ (measured in units of the LJ energy $\epsilon$).
%
We integrate the ensuing equations of motion within a $x$-periodic box of
size $L=N_{\rm p} a_0$, by means of a standard fourth-order Runge-Kutta
method \cite{NumericalRecipes}.
We note that the Nos\'e-Hoover thermostat is not generally well defined for
a forced system in dynamical conditions.  However it can be assumed to work
at least approximately for an adiabatically moving system, where the Joule
heat is a small quantity \cite{Evans85}.

We usually start off the dynamics (for a single lubricant layer) from
equally-spaced lubricant particles at height $r_{i\,z}=a_{\rm b}$ and with
the top layer at height $r_{{\rm t}\,z}= a_{\rm b}+a_{\rm t}$, but we
considered also different initial conditions: after an initial transient,
sometime extending for several hundred time units, the sliding system
reaches its dynamical stationary state.
For many layers we start off with lubricant particles at perfect
triangular lattice sites, and the top slider correspondingly raised.
In the numerical simulations, adiabatic variation of the external driving
velocity is considered and realized by changing $v_{\rm ext}$ in small
steps, letting the system evolve at each step for a time long enough for
all transient stresses to relax.
We compute accurate time-averages of the physical quantities of interest by
averaging over a simulation time in excess of a thousand time units,
starting after the transient is over.
At higher temperature, fluctuations of all physical quantities around their
mean values increase, thus requiring even longer simulation times to obtain
well-converged averages.

\section{The plateau dynamics}\label{plateaudynamics:sec}
We study here the model introduced in Sect.~\ref{model:sec}, firstly for a
single lubricant layer and then for a thicker multilayer of $N_{\rm
  layer}=2\dots\ 8$.
In all cases, we consider complete layers, realizing an essentially
crystalline configuration at the given temperature assumed well below the
melting temperature.
We focus our attention on the the dragging of kinks and on the ensuing
exact velocity-quantization phenomenon.
We expect that, like in previous studies of the idealized 1D model
\cite{Vanossi06,Manini07PRE,Manini07extended,Santoro06,Cesaratto07,Vanossi07Hyst,Vanossi07PRL,Vanossi08TribInt,Manini08Erice},
the ratio $w=\langle v_{\rm cm}\rangle /v_{\rm ext}$ of the
lubricant CM $x$ velocity to the externally imposed sliding speed
$v_{\rm ext}$ should stay pinned to an exact geometrically
determined plateau value, while the model parameters, such as
$v_{\rm ext}$ itself or temperature $T$ or load $F$, are made vary
over wide ranges.
In detail, the plateau velocity ratio
\begin{equation}\label{wplat}
w_{\rm plat}=
\frac{\langle v_{\rm cm}\rangle}{v_{\rm ext}}=
\frac{\frac 1{a_0} - \frac 1{a_{\rm b}} }{\frac 1{a_0}}=
\frac{\lambda_{\rm b} -1 } {\lambda_{\rm b}}=
1-\frac{1}{\lambda_{\rm b}}
\end{equation}
is a function uniquely of the kink linear density, determined by the excess
linear density of lubricant atoms with respect to that of the bottom
substrate, thus of the length ratio $\lambda_{\rm b}$, see
Eq.~(\ref{ratiotb}).
The ratio $\frac{a_0^{-1} - a_{\rm b}^{-1} }{a_0^{-1}}$ represents
precisely the fraction $N_{\rm kink}/(N_{\rm p}/N_{\rm layer})$ of
kink defects in each lubricant layer.
The top length ratio
$\lambda_{\rm t}$, assumed much more different from 1, plays a different
but crucial role, since it sets the
kink coverage $\Theta= N_{\rm kink}/N_{\rm t}=
\left(1-\lambda_{\rm b}^{-1}\right) \lambda_{\rm t}$.
Assuming
that the 1D mapping to the FK model sketched in
Ref.~\cite{Vanossi07PRL} makes sense also in the present richer
geometry, the coverage ratio $\Theta$ should affect the pinning
strength of kinks to the top corrugation, thus the robustness of
the velocity plateau.
We shall try to find out if $\Theta$ assumes a similar role in the 2D model
in Sec.~\ref{doublelayers:sec} below.

In the present work we consider mainly a geometry of nearly full
commensuration of the lubricant to the bottom substrate, $\lambda_{\rm b}$
near unity: in particular we set $\lambda_{\rm b}=29/25=1.16$, which
produces merely $4$ kinks every $29$ lubricant particles in each layer.
This value of $\lambda_{\rm b}$ produces a good kink visibility, but it is
not in any sense special.
We also investigate the plateau dynamics for an anti-kink configuration
$\lambda_{\rm b}=21/25=0.84$.
Even for a $\lambda_{\rm b}$ value significantly distinct from unity, such
as the golden mean $(1+\sqrt 5)/2\simeq 1.62$, we have evidence of perfect
plateau sliding.
The present model allows us to address for the first time the nontrivial
issue of the survival of the quantized plateau even for a somewhat more
realistic 2D dynamics, for several interposed lubricant layers and for
finite temperature.

\subsection{Single lubricant layer }\label{monolayer:sec}

\begin{figure}
\centerline{
\epsfig{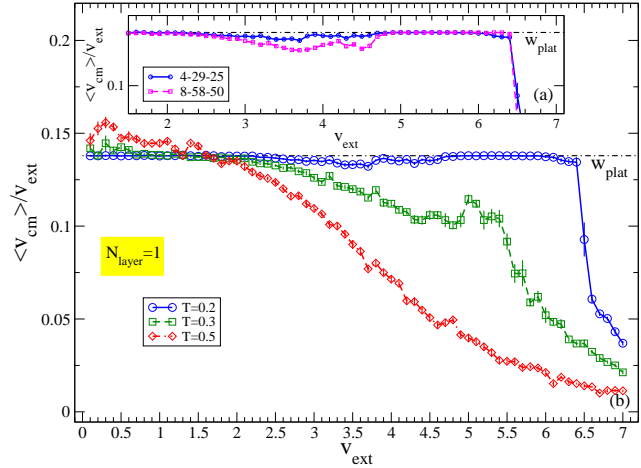} }
\caption{\label{vextmonotheta1:fig}
Average velocity ratio $w=\langle v_{\rm cm}\rangle /v_{\rm ext}$
of a single lubricant layer as a function of the top-layer sliding
velocity $v_{\rm ext}$ (increased adiabatically) for three
different temperatures.
Thermal effects degrade the perfect quantized-velocity plateau which is
very clear at low temperature.
Inset (a): negligible simulation-size dependency of the dynamical critical
depinning point. $4-29-25$ and $8-58-50$ indicate the numbers of particles
$N_t-N_p-N_b$ in the simulations.
All simulations are carried out with $F=25$.
The plateau velocity ratio (dot dashed) is $w_{\rm plat}=\frac{4}{29}\simeq
0.13793$, Eq.~(\ref{wplat}).
}
\end{figure}

Figure \ref{vextmonotheta1:fig} reports the time-averaged
horizontal velocity $\langle v_{\rm cm}\rangle$ of the
single-layer lubricant CM, as a function of the velocity $v_{\rm
ext}$ of a fully commensurate top layer ($\Theta=1$) for
three different temperatures of the system.
The velocity ratio $w=\langle v_{\rm cm}\rangle /v_{\rm ext}$ is generally
a nontrivial function of $v_{\rm ext}$, showing wide flat plateaus and
regimes of continuous evolution.
The plateau velocity matches perfectly the ratio $w_{\rm plat}$ of
Eq.~(\ref{wplat}).
The plateau extends over a wide range of external driving velocities, up to
a critical velocity $v_{\rm crit}$, whose precise value is obtained by
ramping $v_{\rm ext}$ adiabatically; beyond $v_{\rm crit}$, the lubricant
leaves the plateau speed and tends to become pinned to the (better matched)
bottom layer.
On the small-$v_{\rm ext}$ side of the plateau, despite error bars
indicating increasing uncertainty in the determination of
$w$, data are consistent with a plateau dynamics extending all the
way to the static limit $v_{\rm ext}\rightarrow 0$, like in the 1D
model \cite{Manini07PRE}.
As temperature increases, $\langle v_{\rm cm}\rangle$ tends to deviate
slightly from the perfect plateau value.
At the highest temperature considered, $k_BT=0.5\,\epsilon$, near melting of
LJ solid at zero pressure \cite{Ranganathan92}, no plateau is seen in the simulations.
Finite-size scaling, Fig.~\ref{vextmonotheta1:fig}(a), shows
little size effect on the plateau, and in particular on its
boundary edge $v_{\rm crit}$.

\begin{figure}
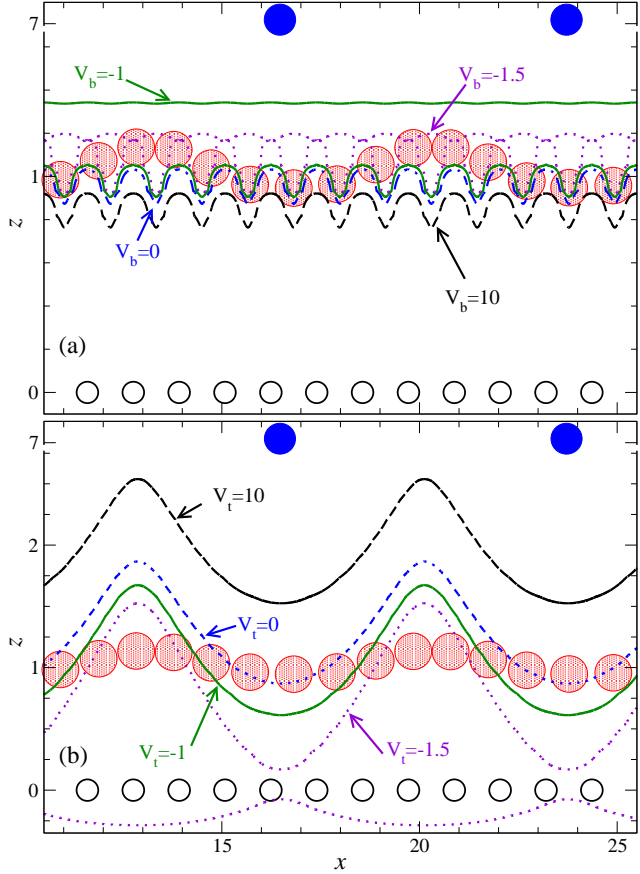

\begin{center}
\epsfig{file=equibottom.eps,width=83mm,angle=0,clip=}\\
\epsfig{file=equitop.eps,width=83mm,angle=0,clip=}
\end{center}
\caption{\label{equi:fig}
Typical positions of top (filled circles), lubricant (shadowed),
and bottom (open) atoms, in the plateau state of a $v_{\rm
ext}=0.1$ $T=0.001$ simulation for one lubricant layer.
Kinks are visible as touching circles.
(a): iso-curves of the potential energy experienced by a single
lubricant particle and produced by the bottom chain; (b):
iso-curves for the top chain.
The equipotential surfaces drawn are $V=10$ (long-dashed), $V=0$
(fine-dashed), $V=-1$ (solid) and $V=-1.5$ (dotted).
The vertical displacements of the lubricant are produced by the
bottom layer pushing kinks out, and enhanced by the interaction
with the perfectly matching ($\Theta=1$) top layer pressed against
the lubricant by a load $F=25$.
}
\end{figure}

The specific roles of the two substrates in the dynamical plateau state is
illustrated by a snapshot of the plateau-state atomic coordinates and
potentials at an arbitrary time, shown in Fig.~\ref{equi:fig}.
The bottom layer produces a potential energy whose iso-levels are sketched
in the upper panel of Fig.~\ref{equi:fig}: with its near-matching
corrugation, this potential profile is responsible for the creation of
kinks, like in the simple 1D model \cite{Vanossi06,Manini07PRE}.
A kink is visible as a local compression of the lubricant atoms
trapped in the same minimum of the bottom substrate potential.
In the quantized-velocity state, kinks pin to the minima of the top
potential (and slide with it at $v_{\rm ext}$), as illustrated in the lower
panel of Fig.~\ref{equi:fig}.

We observe precise velocity quantization also as the downward load $F$
applied to the top layer is changed in magnitude.
At larger temperature, where thermal fluctuations tend to destabilize the
quantized velocity, calculations show that the quantized-velocity state
benefits higher loads $F$.

\subsection{Two lubricant layers and role of kink coverage}
\label{doublelayers:sec}

We now repeat the simulations of the previous Section by considering a doubled
number of lubricant particles in the same box size.
Even when starting from arbitrary geometries, the lubricant atoms
eventually arrange themselves in a regular double layer, a stripe of a
triangular lattice.
After a transient, a quantized-velocity plateau develops, showing
essentially the same conditions as described for a single layer in
Fig.~\ref{vextmonotheta1:fig}, with a clear depinning transition at a
critical velocity $v_{\rm crit}$ remarkably close to that of a single
layer.
In this plateau state, we can still identify kinks in the lubricant layer
adjacent to the bottom potential, while the other layer shows weaker
$x$-spacing modulations.
The vertical displacements of both layers are induced by the interactions
with both the top- and the bottom-layer atoms.

\begin{figure}
\centerline{
\epsfig{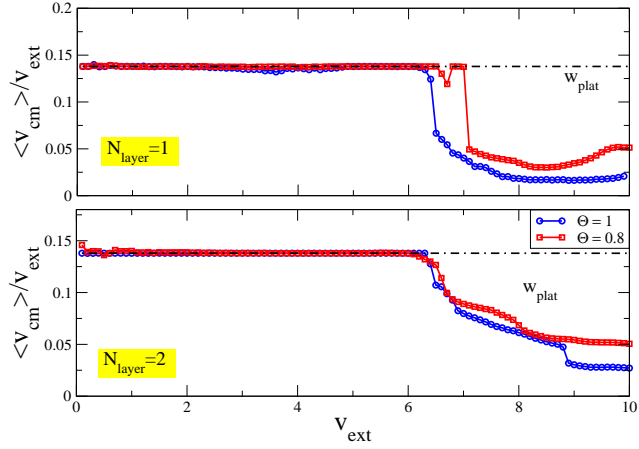} }
\caption{\label{vextcrit:fig}
Comparison of the velocity ratio $w=\langle v_{\rm cm}\rangle
/v_{\rm ext}$ of a lubricant mono-layer and bi-layer as the
top-layer velocity $v_{\rm ext}$ is increased adiabatically, for
kink coverages $\Theta=1$ and $\Theta=0.8$.
All simulations are carried out with $F=25$, $T=0.2$, and
$\lambda_{\rm b}=\frac{29}{25}$ (plateau velocity ratio $w_{\rm
plat}=\frac{4}{29}$).
The velocity $v_{\rm crit}$ at which the plateau dynamics ends
does depend on $\Theta$.
}
\end{figure}
The matching of the number of kinks to the number of top-atoms
$\Theta=N_{\rm kink}/N_{\rm t}=1$ is clearly very favorable for kink
dragging, thus for the plateau phenomenon.
It is important to investigate situations where this strong
commensuration is missing.
As an example of lesser commensuration, we consider $5$, rather than $4$,
particles in the top chain, thus producing a coverage ratio $\Theta =
\frac{4}{5}=0.8$, still commensurate, but weakly so.
Figure~\ref{vextcrit:fig} shows that a perfect plateau again occurs also
for $\Theta=\frac{4}{5}$, whether the lubricant is a mono- or a bi-layer,
and apparently this less commensurate configuration produces and even more
robust quantized-velocity state, at least for $N_{\rm layer}=1$.
%
We note however that this increased stability may be an artifact of having
increased the total load $N_t F$, and thus the applied ``pressure'' on the
lubricant.

\begin{figure}
\centerline{
\epsfig{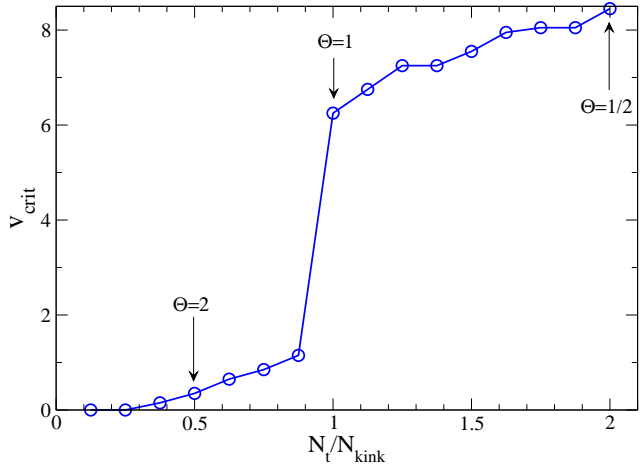}
}
\caption{\label{vcrit:fig}
%
Variation of the plateau boundary velocity $v_{\rm crit}$ as a function of
the inverse coverage $\Theta^{-1}=N_{\rm t}/N_{\rm kink}$ for a lubricant
mono-layer. The calculations show a sudden increase of $v_{\rm crit}$ at
$\Theta^{-1} \geq 1$.
Simulations are carried out with $F=25$, $T=0.2$, and
$\lambda_{\rm b}=\frac{29}{25}$.
}
\end{figure}

It is instructive to study how the depinning point $v_{\rm crit}$ varies
when the ratio of commensuration $\Theta$ varies.
We study this evolution at fixed $\lambda_{\rm b}$, thus fixed
density of solitons, while the number of surface atoms changes in
the top substrate.
Figure~\ref{vcrit:fig} reports the depinning velocity $v_{\rm crit}$,
always evaluated through an adiabatic increase of $v_{\rm ext}$, as a
function of the number $N_{\rm t}$ of top-layer atoms, or the inverse
commensuration ratio $1/\Theta=N_{\rm t}/N_{\rm kink}$.
One mono-layer shows a monotonically increasing depinning velocity $v_{\rm
  crit}$, characterized by a sudden increase in correspondence to the fully
matching coverage $\Theta=1$.
For even larger $N_{\rm t}\sim N_{\rm b}$ (not shown), eventually kinks
cannot ingrain in the much finer oscillation of the top potential energy
and we find a weakening of the plateau regime.
%
%
%
%

\subsection{Lubricant multi-layer}\label{Multilayers:sec}

\begin{figure}
\centerline{
\epsfig{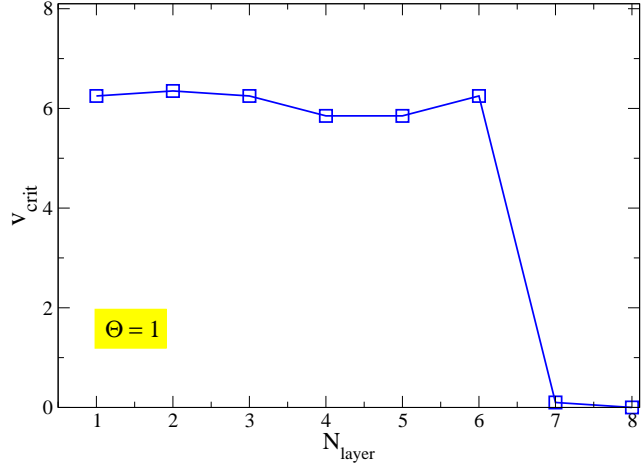}
} \caption{\label{multi:fig}
Critical depinning velocity $v_{\rm crit}$ as a function of the
numbers $N_{\rm layer}$ of lubricant layers, in a condition that
favors pinning: $F=25$, $T=0.2$, $\Theta=1$.
The data show the tendency for $v_{\rm crit}$ to drop considerably
as the number of lubricant layers increases, $N_{\rm layer}>6$,
beyond the boundary-lubricating regime.
}
\end{figure}

\begin{figure}
\centerline{
\epsfig{file=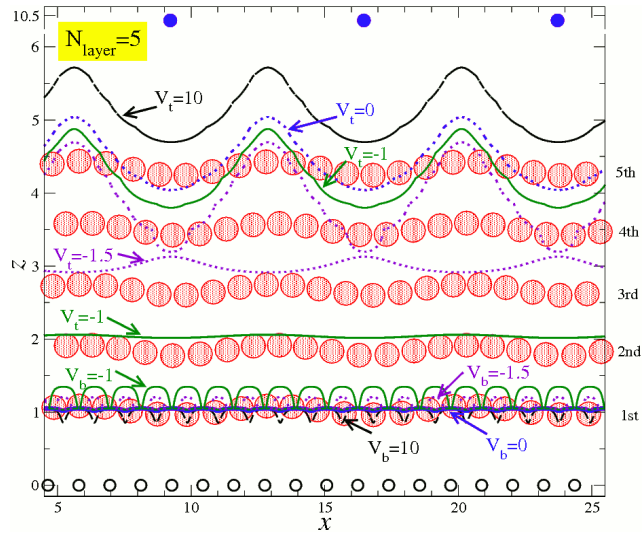,width=83mm,angle=0,clip=} }
\caption{\label{equimulti5:fig} Top (filled circles), lubricant
(shadowed) and bottom (open) atoms in a typical snapshot of the
plateau dynamical state for $N_{\rm layer}=5$: kinks are seen only
in the horizontal displacements of the two lowest layers, while
vertical undulations of all lubricant layers are apparent.
Equal-potential profiles for the potentials created by the top and the
bottom layers, at $V=10$ (long-dashed), $V=0$ (dashed), $V=-1$ (solid) and
$V=-1.5$ (dotted).
All simulations are carried out with $v_{\rm ext}=0.1$, $T=0.001$
and $F=25$.
}
\end{figure}

\begin{figure}
\centerline{ \epsfig{file=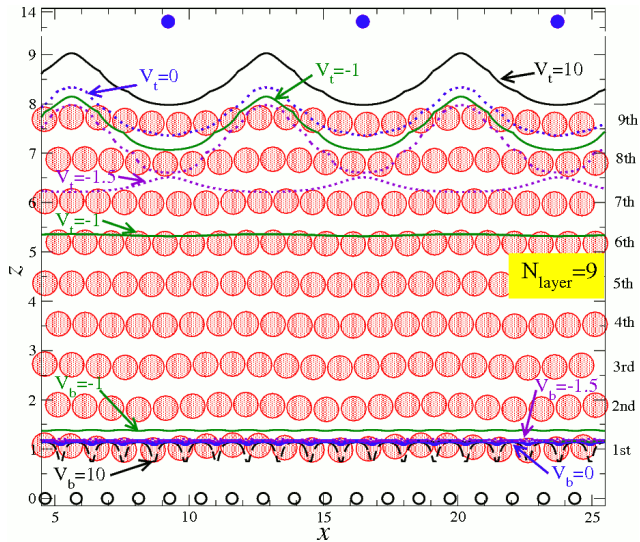,width=83mm,angle=0,clip=}
} \caption{\label{equimulti9:fig} Top (solid circles), lubricant
(shadowed) and bottom (open) atoms in a typical snapshot of the
fragile plateau dynamical state for $N_{\rm layer}=9$: kinks are
seen only in the horizontal displacements of the first two layers,
and vertical spatial modulations persist into the ``thick''
lubricant for about four layers.
Equal-potential surfaces for the potentials created by the top and the
bottom chain lubricant layers at $V=10$ (long-dashed), $V=0$ (dashed),
$V=-1$ (solid) and $V=-1.5$ (dotted).
Remaining simulation parameters are $v_{\rm ext}=0.1$, $T=0.001$
and $F=25$.
}
\end{figure}

We come now to investigate the role of $N_{\rm layer}$ on the dynamically
pinned state.
Figure~\ref{multi:fig} shows the dependency of the critical velocity
$v_{\rm crit}$ on the number $N_{\rm layer}$ of layers of the confined
lubricant in the strong-pinning condition characterized by $\Theta=1$,
$F=25$, and $T=0.001$.
%
For up to $N_{\rm layer}=6$ layers we find quite robust perfect velocity
plateaus, with a remarkably weak dependence of $v_{\rm crit}$ on $N_{\rm
layer}$.
Figure~\ref{equimulti5:fig} shows that little or no sign of kinks
(horizontal displacements) is
visible above the two lowermost layers near the bottom.
However, vertical corrugations of the lubricant propagate from bottom to
top, corresponding to kinks.
These vertical displacements are  mediating agents transmitting the kink
tendency to pin to the top-layer corrugations, and giving rise to the
observed perfect velocity quantization, at least for small $v_{\rm ext}$.

For a further increase in $N_{\rm layer}$, this $z$-displacement
mechanism becomes rapidly ineffective, as evident in
Fig.~\ref{equimulti9:fig}: the vertical corrugations induced by
the substrates reach into the solid lubricant for about $4$
layers, while inner layers, such as the $5$th layer of
Fig.~\ref{equimulti9:fig} remain essentially flat, thus not
supporting the dynamic pinning.
In the unpinned state, the top-chain slides over the upper lubricant layer,
but the deformation it induces propagates only through a few superficial
layers but cannot drag the kinks created by the bottom potential.
Even in the large-$v_{\rm ext}$ unpinned state, the relative
positions of lubricant atoms are essentially ordered, and show
neither defects nor a liquid configuration, due to the low
temperature considered, confinement \cite{Klein98,Persson99}, and
full commensuration.
%

\subsection{Anti-kinks}\label{antikink:sec}

\begin{figure}
\centerline{
\epsfig{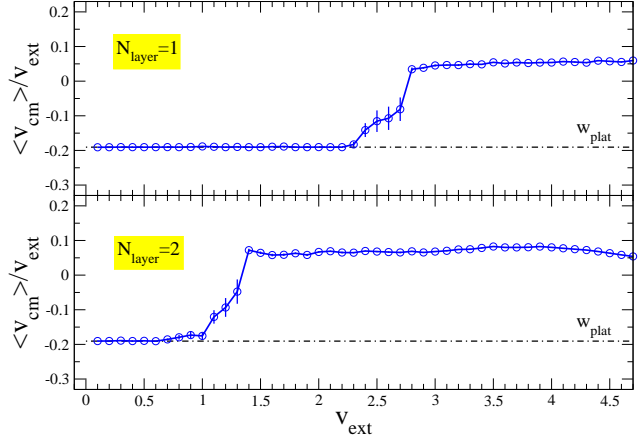}
}
\caption{\label{antikink:fig}
Average velocity ratio $w=\langle v_{\rm cm}\rangle /v_{\rm ext}$ as a
function of the top-layer velocity $v_{\rm ext}$ (increased adiabatically)
for a model composed by $4$, $21$ and $25$ atoms in the top, lubricant and
bottom chains: these correspond to a $\lambda_{\rm b}=21/25=0.84$, which
according to Eq.~(\ref{wplat}), produces perfectly quantized dynamics at
$w_{\rm plat}=-\frac{4}{21} \simeq -0.190476$ (dot-dashed line).
The other simulation parameters are: $F=25$, $T=0.2$, and
$\Theta=1$.
}
\end{figure}

Previous 1D work showed the surprising phenomenon of {\em backward}
lubricant sliding corresponding to the dragging of anti-kinks
\cite{Vanossi06,Manini07PRE}.
We set now a reversed condition of quasi-commensuration of the chain to the
$a_{\rm b}$ substrate, $\lambda_{\rm b}=21/25=0.84$, which produces a {\em
negative} $x$-density of kinks $-4/21\,a_0^{-1}\simeq-0.190\,a_0^{-1}$.
This condition in fact produces, instead of local compressions, local
dilations of the chain, classifiable as {\em anti-kinks}, alternating with
in-register regions.
The anti-kinks again pin, like kinks did, to the corrugations of the top
substrate, which drag them along at full speed $v_{\rm ext}$.
As anti-kinks are basically missing particles, like holes in
semiconductors, they carry a negative mass.
Their rightward motion produces a net leftward motion of the lubricant: the
lubricant chain moves in the {\em opposite direction} with respect to the
top layer \cite{Vanossi06}.
Figure~\ref{antikink:fig} displays a clear reversed-velocity plateau for
both one layer and two layers, thus confirming this mechanism.
The perfect plateau is comparably weaker than the plateau produced by
$\lambda_{\rm b}>1$, as seen from it ending at a smaller $v_{\rm ext}$.

\section{Discussion and conclusions}
Within the idealized scheme of a simple 1D FK-like model, a special
``quantized'' sliding state was found for a solid lubricant confined
between two periodic layers \cite{Manini07extended}.
This state, characterized by a nontrivial geometrically fixed ratio of the
mean lubricant drift velocity $\langle v_{\rm cm}\rangle$ and the
externally imposed translational velocity $v_{\rm ext}$, was understood as
due to the rigid dragging of kinks (or solitons), formed by the lubricant
due to incommensuration with one of the substrates, pinning to the other
sliding substrate.
In the present work, a quantized sliding state of the same nature is
demonstrated for a substantially less idealized 2D model of boundary
lubrication, where atoms are allowed to move perpendicularly to the sliding
direction and interact via LJ potentials.
We find perfect plateaus, at the same geometrically determined velocity
ratio $w_{\rm plat}$ as observed in the simple 1D model for varied driving
speed $v_{\rm ext}$, not only at low temperatures but also for temperatures
not too far from the melting point of the LJ lubricant, whether the model
solid lubricant runs from a single layer to several layers.
An increased load-$F$ tends to benefit the plateau state at higher
temperatures.
The velocity plateau, as a function of $v_{\rm ext}$, ends at a critical
velocity $v_{\rm crit}$, and for $v_{\rm ext}>v_{\rm crit}$ the lubricant
moves at a speed which is generally lower than that of the plateau
state.
In fact, by cycling $v_{\rm ext}$, the layer sliding velocity exhibits
hysteretic phenomena around $v_{\rm crit}$, which we shall investigate in
detail in future work.
The unpinning velocity $v_{\rm crit}$ is linked to the
commensuration $\Theta$ of kinks to the upper slider period: at
$\Theta=1$ marks a sudden rise of $v_{\rm crit}$.
%
A clear plateau dynamics is demonstrated even for a confined solid
lubricant composed of several (up to $N_{\rm layer}=6$) lubricant layers:
the strength of the plateau (measured by $v_{\rm crit}$) is a generally
decreasing function of the number of layers.
The striking backward lubricant motion produced by the presence of
``anti-kinks'' is again recovered in this more realistic context.
The present work focuses on ordered configurations: both substrates
are perfect crystals and the lubricant retains the configuration
of a strained crystalline solid.
The dynamical depinning speed $v_{\rm crit}$, that we usually find of the
order of a few model units (corresponding to $\sim 10^3$~m/s for realistic
choice of the model parameters), is very large compared to typical sliding
velocities investigated in experiments.  This suggests that sliding at a
dynamically quantized velocity is likely to be extremely robust.  In
experiments, depinning from the quantized sliding state is likely to be
associated to mechanisms such as disorder or boundary effects, rather than
excessive driving speed.
The role of disorder and defects both in the substrate \cite{Guerra07} and
in the lubricant will be the object of future investigation.
A detailed investigation of the stick-slip phenomena and of other features
of the dynamical properties will also require further study.

\section*{Acknowledgments}

This research was partially supported by PRRIITT (Regione Emilia
Romagna), Net-Lab ``Surfaces \& Coatings for Advanced Mechanics
and Nanomechanics'' (SUP\&RMAN).
Work in SISSA was supported through PRIN 2006022847, and Iniziativa
Trasversale Calcolo Parallelo INFM-CNR.


\section*{References}


\begin{thebibliography}{10}

\bibitem{Vanossi06}
{ A.\ Vanossi, N.\ Manini, G.\ Divitini, G.\ E.\ Santoro, and E.\ Tosatti,
  Phys.\ Rev. Lett.\ {\bf 97}, 056101 (2006)}.

\bibitem{Manini07PRE}
{ N.\ Manini, A.\ Vanossi, G.\ E.\ Santoro, and E.\ Tosatti, Phys.\ Rev. E {\bf
  76}, 046603 (2007)}.

\bibitem{nose-hoover}
{ D.\ Frenkel and B.\ Smit, {\it Understanding Molecular Simulation.\ From
  Algorithms to Applications} (Academic Press, London, 1996)}.

\bibitem{Martyna92}
{ G.\ J.\ Martyna, M.\ L.\ Klein and M.\ Tuckerman, J.\ Chem.\ Phys. {\bf 97},
  2635 (1992)}.

\bibitem{NumericalRecipes}
{ W.\ H.\ Press, S.\ A.\ Teukolsky, W.\ T.\ Vetterling and B.\ P.\ Flannery,
  {\it Numerical Recipes in Fortran.\ The Art of Parallel Scientific Computing}
  (Cambridge University Press, Cambridge, 1996)}.

\bibitem{Evans85}
{ D.\ J.\ Evans and B.\ L.\ Holian, J.\ Chem.\ Phys. {\bf 83}, 4069 (1985);}.

\bibitem{Manini07extended}
{ N.\ Manini, M.\ Cesaratto, G.\ E.\ Santoro, E.\ Tosatti, and A.\ Vanossi, J.\
  Phys.: Condens.\ Matter {\bf 19}, 305016 (2007)}.

\bibitem{Santoro06}
{ G.\ E.\ Santoro, A.\ Vanossi, N.\ Manini, G.\ Divitini, and E.\ Tosatti,
  Surf.\ Sci. {\bf 600}, 2726 (2006)}.

\bibitem{Cesaratto07}
{ M.\ Cesaratto, N.\ Manini, A.\ Vanossi, E.\ Tosatti, and G.\ E.\ Santoro,
  Surf.\ Sci. {\bf 601}, 3682 (2007)}.

\bibitem{Vanossi07Hyst}
{ A.\ Vanossi, G.\ E.\ Santoro, N.\ Manini, M.\ Cesaratto, and E.\ Tosatti,
  Surf.\ Sci. {\bf 601}, 3670 (2007)}.

\bibitem{Vanossi07PRL}
{ A.\ Vanossi, N.\ Manini, F.\ Caruso, G.\ E.\ Santoro, and E.\ Tosatti, Phys.\
  Rev. Lett.\ {\bf 99}, 206101 (2007)}.

\bibitem{Vanossi08TribInt}
{ A.\ Vanossi, G.\ E.\ Santoro, N.\ Manini, E.\ Tosatti, and O.\ M.\ Braun,
  arXiv:0709.4374 [cond-mat.mtrl-sci], submitted to Tribology International}.

\bibitem{Manini08Erice}
{ N.\ Manini, G.\ E.\ Santoro, E.\ Tosatti, and A.\ Vanossi, in print in J.\
  Phys.: Condens.\ Matter (2008)}.

\bibitem{Ranganathan92}
{ S.\ Ranganathan and K.\ N.\ Pathak, Phys.\ Rev. A {\bf 45}, 5789 (1992)}.

\bibitem{Klein98}
{ J.\ Klein and E.\ Kumacheva, J.\ Chem.\ Phys. {\bf 108}, 6996 (1998)}.

\bibitem{Persson99}
{ B.\ N.\ J.\ Persson, Surf.\ Sci. Rep.\ {\bf 33}, 83 (1999)}.

\bibitem{Guerra07}
{ R.\ Guerra, A.\ Vanossi, and M.\ Ferrario, Surf.\ Sci. {\bf 601}, 3676
  (2007)}.

\end{thebibliography}

\end{document}